\documentclass[
]{ceurart}

\usepackage[frozencache=true,cachedir=minted-cache]{minted} 
\usepackage{color}
\usepackage{graphics}

\setminted{breaklines=true}

\begin{document}

\copyrightyear{2021}
\copyrightclause{Copyright for this paper by its authors.
  Use permitted under Creative Commons License Attribution 4.0
  International (CC BY 4.0).}

\conference{CLEF 2021 -- Conference and Labs of the Evaluation Forum, 
	September 21--24, 2021, Bucharest, Romania}

\title{Weakly-Supervised Classification and Detection of Bird Sounds in the Wild. A BirdCLEF 2021 Solution}


\author[1,4]{Marcos V. Conde}[%
email=drmarcosv@protonmail.com,
]
\address[1]{Universidad de Valladolid, Spain}

\author[2,4]{Kumar Shubham}[%
email=kumar.shubham@alumni.iitd.ac.in,
]
\author[3,4]{Prateek	Agnihotri}[%
]

\address[2]{Jio Saavn, India}
\address[3]{Clairvoyant.ai, India}
\author[4]{Nitin D. Movva}[]

\author[]{Szilard	Bessenyei}[]

\address[4]{Equal contribution.}

\begin{abstract}
It is easier to hear birds than see them, however, they still play an essential role in nature and they are excellent indicators of deteriorating environmental quality and pollution. Recent advances in Machine Learning and Convolutional Neural Networks allow us to detect and classify bird sounds, by doing this, we can assist researchers in monitoring the status and trends of bird populations and biodiversity in ecosystems.
We propose a sound detection and classification pipeline for analyzing complex soundscape recordings and identify birdcalls in the background. Our pipeline learns from weak labels, classifies fine-grained bird vocalizations in the wild, and is robust against background sounds (e.g., airplanes, rain, etc).
Our solution achieved 10th place of 816 teams at the BirdCLEF 2021 Challenge hosted on Kaggle.\\
Code and models will be open-sourced at \url{https://github.com/kumar-shubham-ml/kaggle-birdclef-2021}.

\end{abstract}


\maketitle

\section{Introduction}
\label{sec:intro}

The BirdCLEF 2021 Challenge \cite{birdclef2021, lifeclef2021} proposes to identify bird calls in soundscape recordings. The challenge was hosted on Kaggle from April 1, 2021 to June 1, 2021 \footnote{\url{https://www.kaggle.com/c/birdclef-2021/}}.
\paragraph{Dataset.}
The training set consists of short audio recordings of 397 bird species generously uploaded by users of \url{xenocanto.org}. These audio files have been downsampled to 32 kHz and converted to the $ogg$ format. In Section \ref{sec:pre} we explain how we preprocess this short audios and generate curated audios and their corresponding Mel Spectrogram. The test set contains approximately 80 soundscape recordings in $ogg$ format (over 10 minutes of recordings), note that participants cannot access these audios.
Additionally, recordings have associated metadata as the location (longitude, latitude), author, date, etc. Some of this features as the location can be especially useful for identifying migratory birds.
\paragraph{Problem.}
Given a long audio in $ogg$ format, participants have to predict if there is a bird call in each 5-seconds segment of the given soundscape, and identify which of the 397 birds is in such segment, thus, once the call is detected in a segment, the task can be considered as fine-grained multi-label classification.
Models infer on the test set with 3 hours run-time limit, to ensure the efficiency of the solutions. 

\paragraph{Evaluation.}
The performance is measured using the ``micro averaged F1 score'' and reported on a Leaderboard (LB). Moreover, this leaderboard is divided into: ``Public'' which provides the score on 28 test recordings (35\%), and Private, which provides the score on 52 test recordings (65\%). During the competition, the participants only get feedback of their performance from the public leaderboard, this is done to prevent overfitting.\\

Train recordings were uploaded by the users of xenocanto.com from sites across the globe; however, test recordings were from four places only:

\begin{enumerate}
    \item \textbf{COL} Jardín, Departamento de Antioquia, Colombia
    \item \textbf{COR} Alajuela, San Ramón, Costa Rica
    \item \textbf{SNE} Sierra Nevada, California, USA
    \item \textbf{SSW} Sapsucker Woods, Ithaca, New York, USA
\end{enumerate}

\begin{figure}[h!]
    \centering
    \setlength{\tabcolsep}{2.0pt}
    \begin{tabular}{cccc}
    \includegraphics[width=0.25\textwidth]{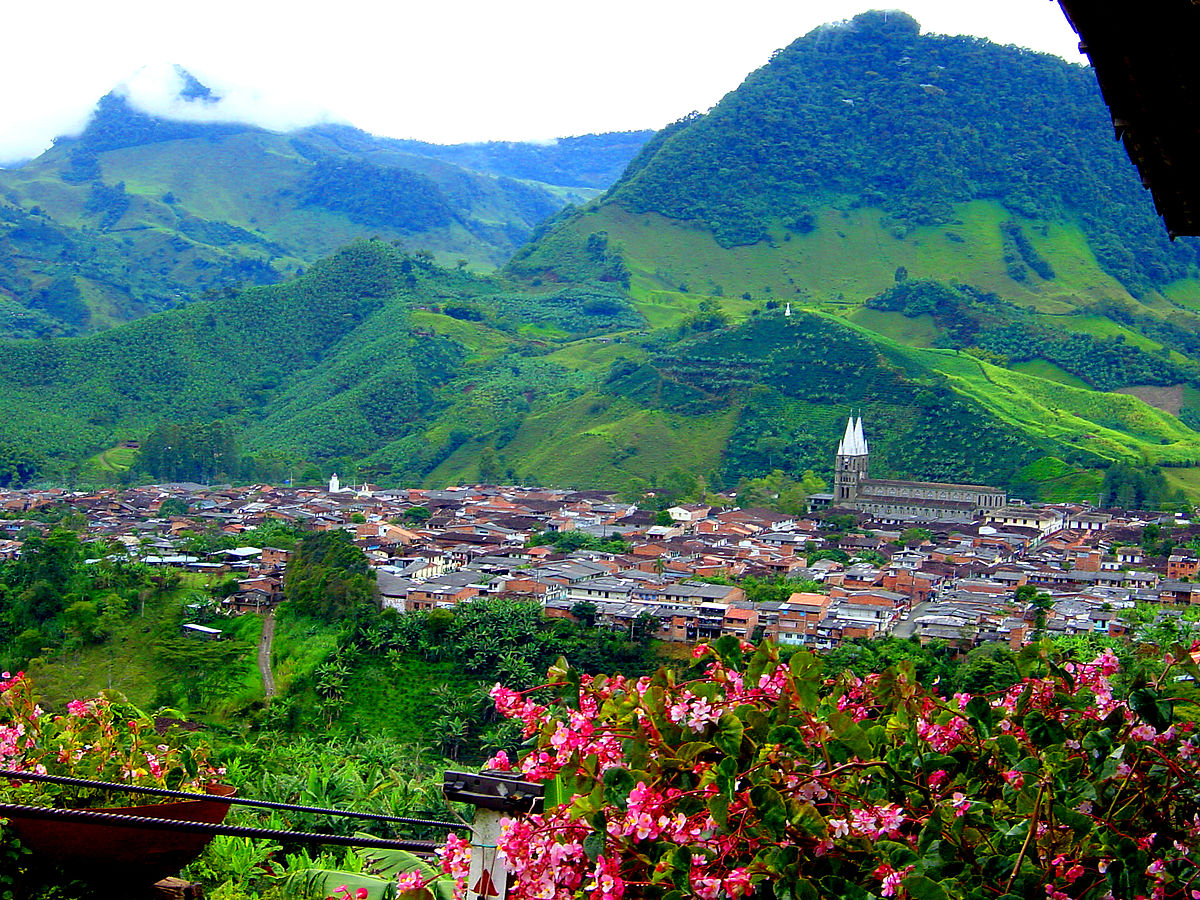} & 
    \includegraphics[width=0.25\textwidth]{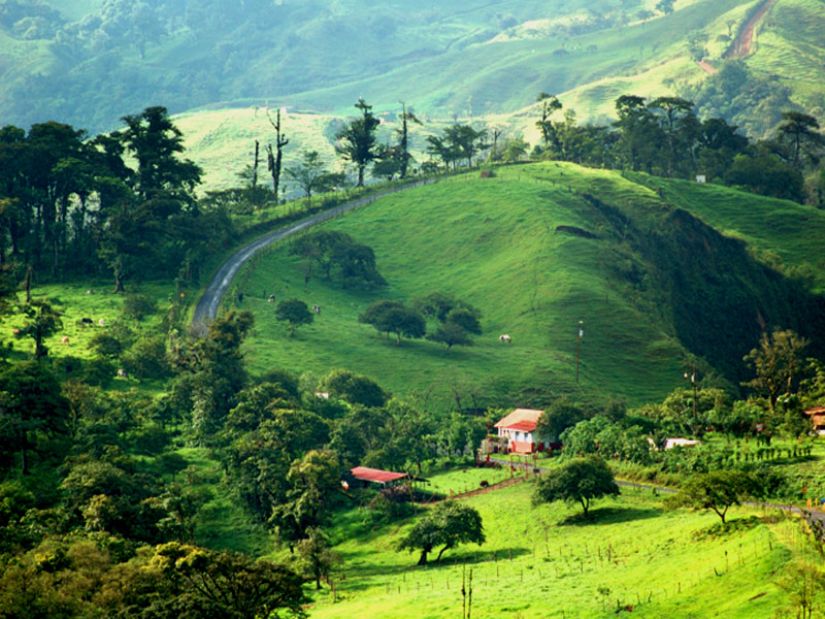} &
    \includegraphics[width=0.25\textwidth]{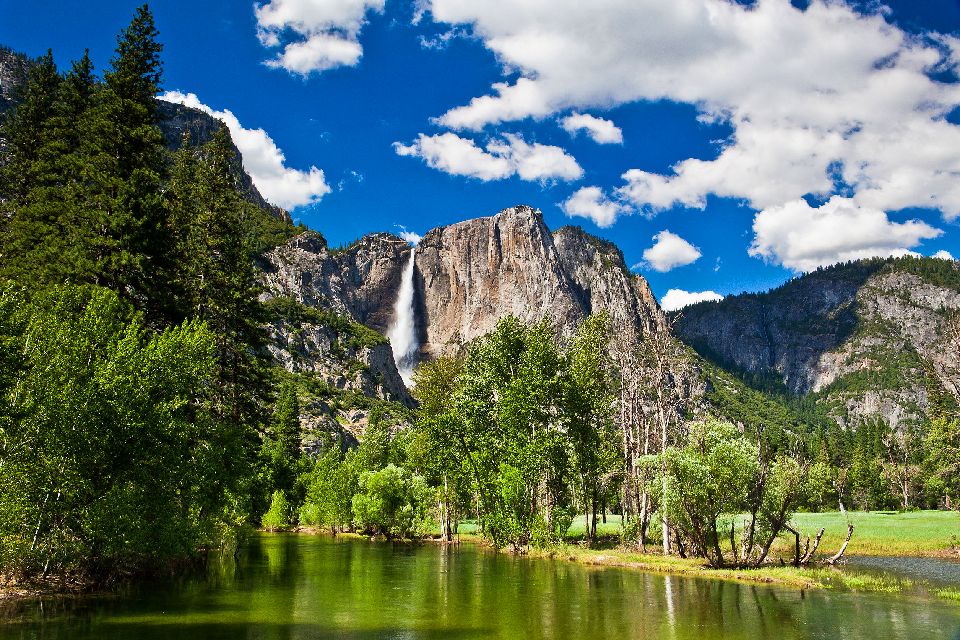} &  \includegraphics[width=0.25\textwidth]{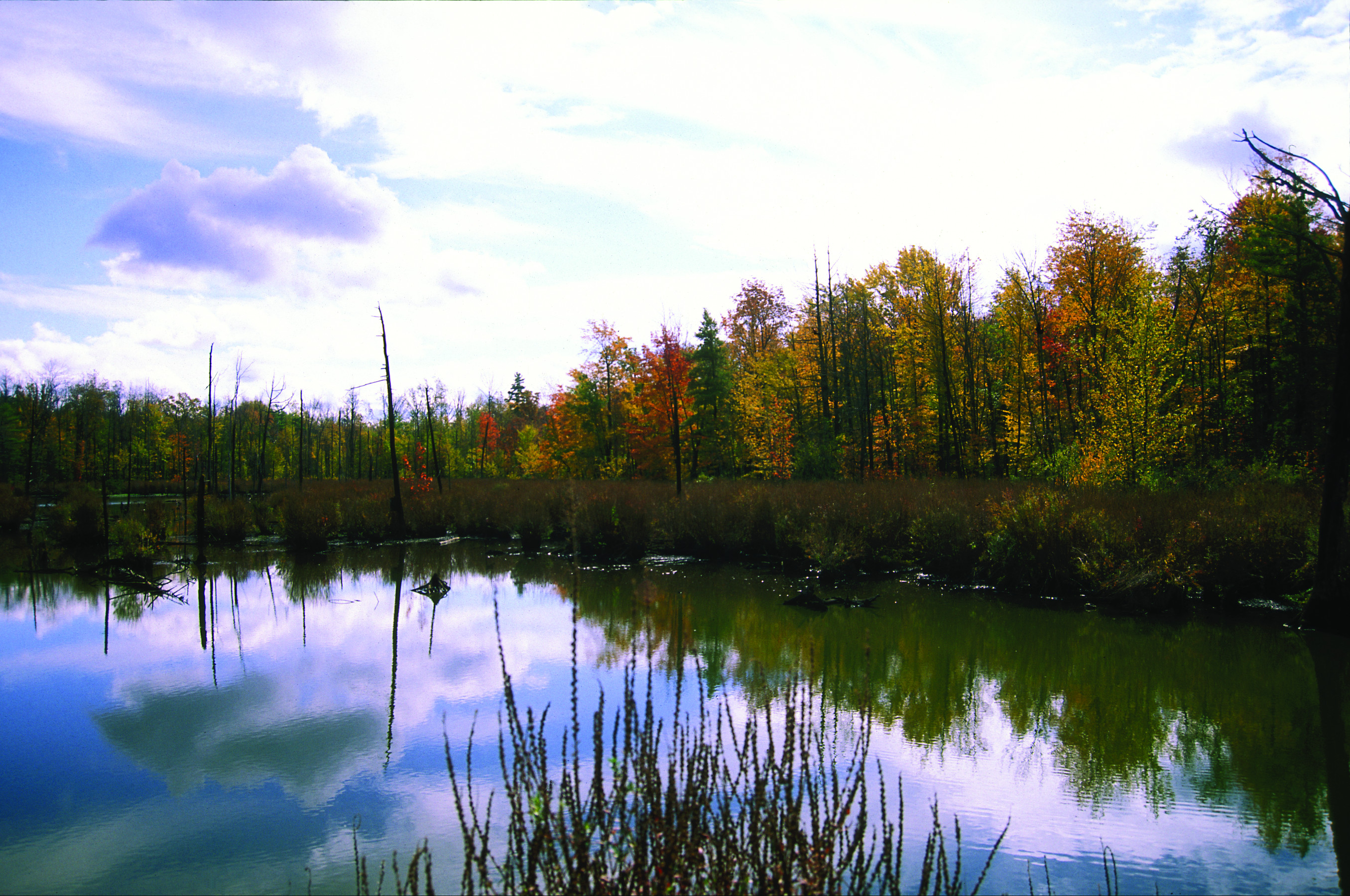}\tabularnewline
    COL & COR & SNE & SSW  \tabularnewline
    \end{tabular}
    \caption{Photographs of the four different sites from which audios were recorded.}
    \label{fig:places}
\end{figure}

We define some terms related with this challenge that we will use in this work: 
\begin{itemize}
    \item Leaderboard denoted as LB (including its two variants, public and private)
    \item Cross-Validation denoted as CV.
    \item the so-called ''score`` or ''metric`` refers to the official challenge metric: ''the row-wise micro averaged F1 score.``.
    \item We refer to the ''Cornell Birdcall Identification - Kaggle 2020`` as the ''previous competition``, ''last year challenge``.
    \item We define ''nocall`` as the class corresponding to the events in an audio where birdcalls are not detected. Other authors might also refer to this term as ''nosound`` or ''background``. This concept might be mentioned various times throughout the manuscript in different notations (“nocall”, nocall, no\_call, etc).
    \item Train soundscapes (or ''train soundscapes``) are 20 audio files that are quite comparable to the test set. They are all roughly ten minutes long and in the $ogg$ format.
\end{itemize}

\section{Related Work}
\label{sec:work}

Previous years BirdCLEF challenges proposed different problems related with large-scale bird recognition in soundscapes or complex acoustic environments \cite{birdclef2021, inproceedings_birds2020, Kahl2019OverviewOB}.
Sprengel~\textit{et.al.} \cite{Sprengel2016AudioBB} and Lasseck \cite{Lasseck2019BirdSI,Lasseck2018AudiobasedBS} introduced deep learning techniques for the ''Bird species identification in soundscapes`` problem.
State-of-the-art solutions are based on Deep Convolutional Neural Networks (CNNs) \cite{Schlter2018BirdIF, Bai2020XceptionBM_birds, Mhling2020BirdSR}, usually, deep CNNs with attention mechanisms are selected as backbone in these experiments \cite{zhang2020resnest, tan2020efficientnet, xie2017aggregated-resnext, kong2020panns, he2015deep}.
Pretrained audio neural networks (PANNs) \cite{kong2020panns} provide a multi-task state-of-the-art baseline for audio related tasks, in previous competitions these networks proved their generalization capability.
Other approaches are focused on Sound Event Detection (SED) \cite{inproceedings_sed2020, fonseca2019learningsed, kong2020panns, article_sed, article_sed}, these approaches usually employ 2D CNNs to extract useful features from the input audio signal (log-melspectrogram), these features still contain information about frequency and time, then recurrent neural networks (RNNs) are used to model longer temporal context from the extracted features or use directly the feature map to predict, since it preserves time segment information.


\section{Proposed Solution}
\label{sec:method}

In this Section we explain the main components of our solution to the BirdCLEF 2021 Birdcall Identification Challenge. We base our solution on diverse and robust models trained on a complete audio dataset using custom augmentations, and on a postprocess algorithm that improves the predicted probabilities of bird appearances by using additional features as the site (longitude, latitude), rarity of the bird, appearance of other birds in the audio, etc.

\begin{figure}[]
    \centering
    \includegraphics[width=\textwidth]{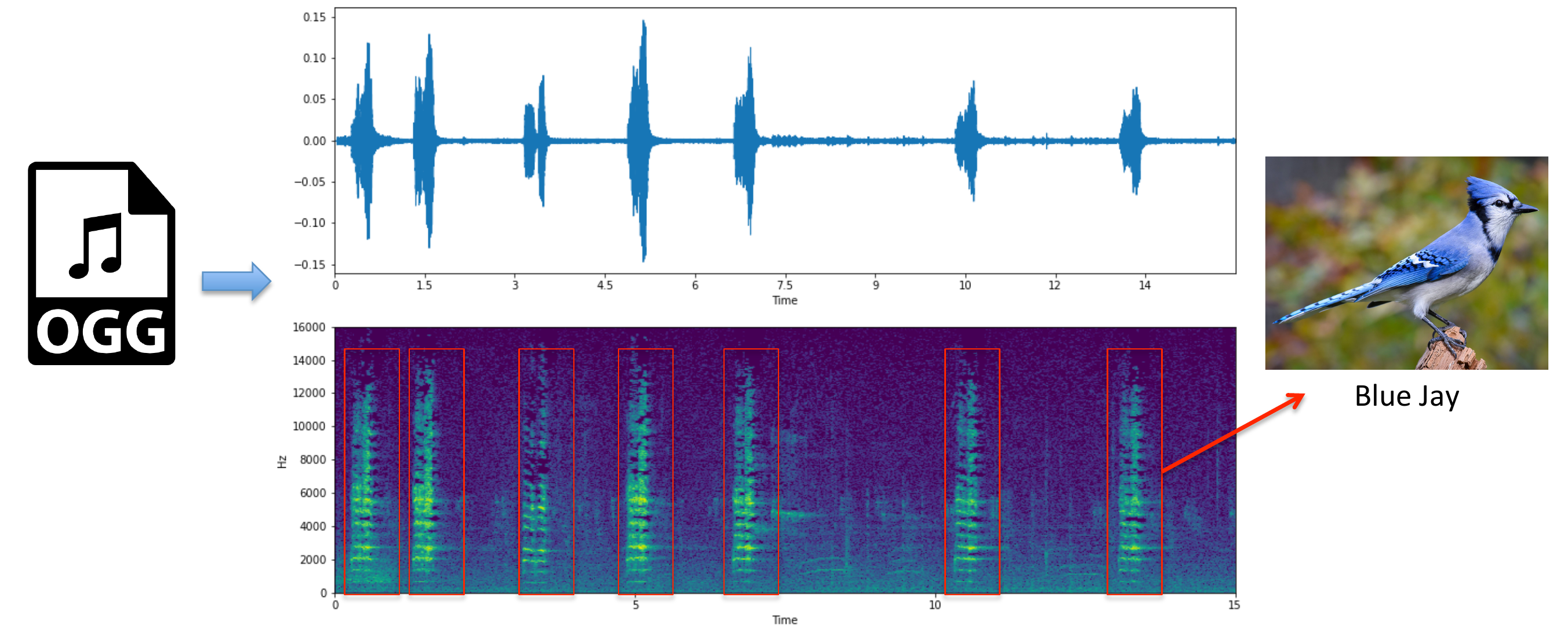}
    \caption{Problem example. The audio in $ogg$ format can be visualized as a waveform (top) or a Mel Spectrogram (bottom). We draw a red bounding box on the events in the audio where a birdcall is detected, the other parts of the audio might contain background sounds or noise, such parts where we do not detect a birdcall are also called ''nocall`` events. We detect the bircall in the audio, and also identify the corresponding bird, in this case a ''Blue Jay`` (Cyanocitta cristata).}
    \label{fig:general}
\end{figure}

\subsection{Dataset Preprocessing}
\label{sec:pre}

We converted all the raw audio data to Mel Spectrograms using the $librosa$ library with each having a length of 7 seconds and having some overlap \footnote{\url{https://www.kaggle.com/kneroma/kkiller-birdclef-2021}}, we use this length instead of 5s to ensure that the birdcalls are present in the clip.
The Spectrograms were generated using the following parameters: 
sample rate 32.000, 128 number of mels, minimum frequency 0 Hz, maximum frequency 16000 Hz, length of fast Fourier transform window (n-fft) 3200, and number of samples between successive frames (hop-length) set to 80. 
We use the Cornell Birdcall Identification 2020 Challenge dataset \footnote{\url{https://www.kaggle.com/c/birdsong-recognition}} as additional data, this dataset has 183 birds in common and allowed us to add 1300 extra audio files.
After some visual inspections of the Mel Spectrograms, we determine a threshold such that the spectogram is considered to have weak a signal or no signal, attending to its mean and maximum values. All the spectrograms with no signal or very weak signals are removed and treated as noise.
Once the above preprocessing steps are completed, we split the training data into 5 different stratified folds. 

\subsection{Augmentations}
\label{sec:augs}

\begin{figure}[!ht]
    \centering
    \includegraphics[width=\textwidth]{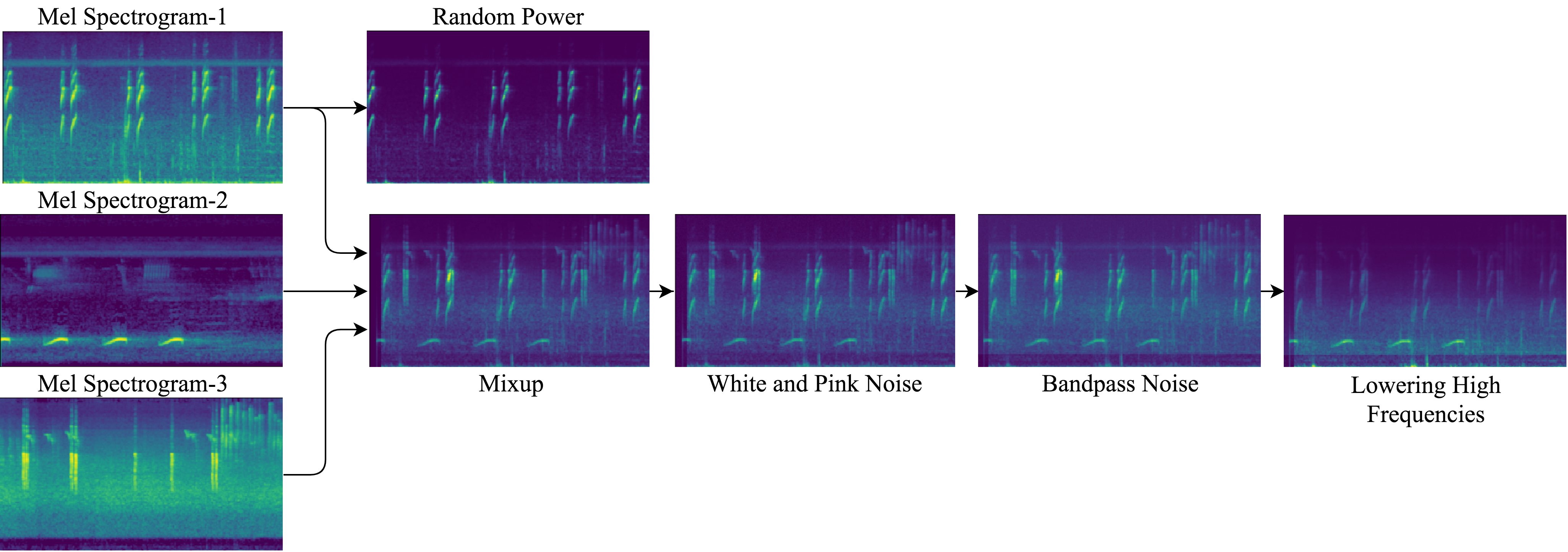}
    \caption{Visualization of our augmentation pipeline explained in Section \ref{sec:augs}.}
    \label{fig:augs}
\end{figure}

We use 6 different types of augmentations in order to improve the robustness and generalization capability of our models. In Figure \ref{fig:augs} we show the effect of the proposed augmentations in the same order we apply them: Mixing of images, Random Power, White noise, Pink Noise, Bandpass noise, Lower the upper frequencies.
%

First, 2 or 3 different training images are overlapped on each other with a random probability of mixing (default is $0.7$). Once this is completed random power is applied on the mixed image to bring all the images to a certain contrast and brightness level. 
Next we add augmentations in the following order: white noise, pink noise, bandpass noise, reducing upper level frequencies, we found experimentally that this is the optimal order.
All the augmentations mentioned above are added with a probability between $0.4$ and $0.7$ to ensure the diversity of the data. 

\subsection{Models}
\label{sec:models}
In Table \ref{tab:archs} we show the model architectures used in our experiments. All the models had similar performance on out-of-fold validation using 5 stratified folds. Single models perform reasonably good, but combining them into an ensemble provided best performance as we explain in Sections \ref{sec:infer} and \ref{sec:exp}.
In our experiments we found that bigger architectures did not necessarily provide better results. Hence, a lot of experimentation was done with smaller models such as ResNeSt-50 \cite{zhang2020resnest} and EfficientNet-B0 \cite{tan2020efficientnet}. 
In addition to the proposed models, we use top models from the Cornell Birdcall Identification 2020 Challenge \footnote{\url{https://www.kaggle.com/c/birdsong-recognition}}, in Section \ref{sec:exp} we explain how we incorporate the following models into our ensemble:

\begin{enumerate}
    \item The 1st place solution there are 14 models in total, all of which are PANN DenseNet-121 architecture with an added attention layer. The models were trained with 264 classes of bird data and augmentations such as SpecAugmentation, gaussian noise, gain (volume adjustment), along with mixup for some models were used to increase model’s robustness.
    
    \item The 2nd place solution. Two different models: ResNet-50 and EfficientNet-B0. Both these architectures were trained with different settings on 264 birds and were trained directly on mel spectrograms instead of training on the audio files.
\end{enumerate}

\begin{figure}[ht!]
    \centering
    \includegraphics[width=\textwidth]{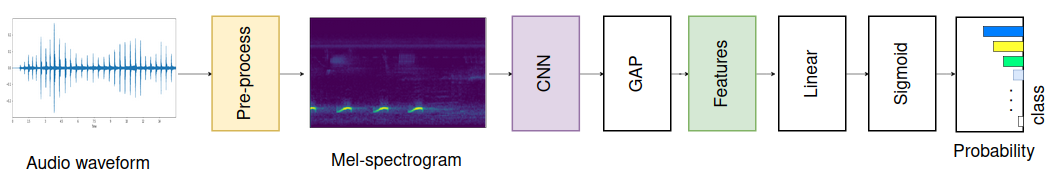}
    \caption{Example of multilabel classification model pipeline. During training, the generated Mel Spectrogram (see Section \ref{sec:pre}) is augmented as explained in Section \ref{sec:augs}. In this diagram we do not show additional postprocess of the predictions.}
    \label{fig:model}
\end{figure}

\subsection{Training Details}

\begin{figure}[h!]
    \centering
    \includegraphics[width=\textwidth]{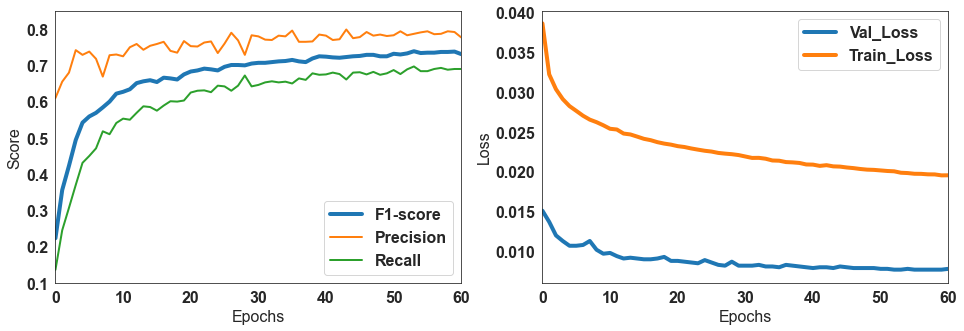}
    \caption{Loss and validation metrics evolution during training of ResNeSt-50 model. Note that the training loss is higher than the validation loss because augmentations were only applied during training.}
    \label{fig:metrics}
\end{figure}

We use a GPU RTX-2070 with 8 GB VRAM for training our models, training time for each model on this device is reported on Table \ref{tab:archs}. 
In all the experiments we train for 60 epochs, we use batch size 64 and Adam Optimizer \cite{kingma2017adam}. We use a Binary Cross Entropy loss function implemented as in PyTorch with Label Smoothing \cite{muller2020doeslabel}.
Additionally we use a Learning Rate Scheduler based on Cosine Annealing with base LR of 0.001 \cite{loshchilov2017sgdr}.
During training we track the loss function, F1 score, Precision, Recall, Label ranking average precision score for both training and validation data. See Figure \ref{fig:metrics} as an example of our training metrics monitoring.

\begin{table*}[]
    \centering
    \caption{Ablation study of our models trained from scratch on the BirdCLEF 2021 dataset. The training time depends on the number of augmentations, architecture and batch size. The corresponding Out-Of-Fold (5-fold) F1 score for each model and the validation score using Train Soundscapes (TS) are provided. Models with ResNeSt as backbone have better performance than DenseNet or EfficientNet.
    } 
    \begin{tabular}{l c c c c}
    \toprule
    Architecture & OOF F1 Score  & TS F1 Score & No. Parameters & Time (min) $\times$ Epoch\\
    \midrule
    ResNeSt-50        \cite{zhang2020resnest} & \textbf{0.755} & \textbf{0.706} & 26,247.693 & 20 \\
    ResNeSt-101       \cite{zhang2020resnest} & 0.748 & 0.705 & 47,039.469 & 34 \\
    ResNeXt-50\_32x4d \cite{xie2017aggregated-resnext}& 0.714 & 0.63 & 23,793.357 & 22 \\
    SeResNet-50       \cite{hu2019squeezeandexcitation} & 0.725 & 0.674 & 26,852.477 & 25 \\
    DenseNet-121      \cite{huang2018densely}& 0.718 & 0.66 & 7,360.781 & 17 \\
    EfficientNet-B0   \cite{tan2020efficientnet} & 0.722 & 0.691 & \textbf{4,516.105} & \textbf{15} \\
    \bottomrule
    \end{tabular}
    \label{tab:archs}
\end{table*}

\subsection{Inference and Postprocessing}
\label{sec:infer}

We use the provided ''train soundscapes`` audio samples as validation set. These audios were much noisier than the curated ones used for training (see Section \ref{sec:pre}) and closely resemble ''test soundscapes``. The distribution of birds in these soundscapes was also different from the training short audios. However, there were only 20 soundscape clips, which covered only 48 of 397 bird classes and 2 of 4 possible sites, making it too challenging to train acoustic models on these clips. The training short audios were only labelled at clip level, but the long audio predictions were generated at frame level (frame of 5 seconds). The noisy labels led to a significant gap between the performance of our models on short audios (reported at Table \ref{tab:archs}) and this validation set. For these reasons, the ''train soundscapes`` clips were used only for validation purposes and to achieve better generalization.

A series of post-processing strategies were employed to bridge the gap between performance on short, cleaner audio and soundscapes. Our post-processing 
improved the cross-validation (CV) and Leaderboard F1 score (LB) by $0.008$-$0.01$. 

Initially, we infer all 5-second clips at a stride of 1 second. Our final submission consists of an ensemble of 13 different models explained at Section \ref{sec:models}. The ensemble optimized weights were calculted based on the validation set. 
A second-stage model, Support Vector Classifier, was trained with a leave-one-clip-out validation strategy on the 20 train soundscapes. This model computes calibrated confidence based on some frame-based, clip-based, and distance-based features generated from probabilities per inference step, in addition to latitude and longitude information for the sites. During the training of the second-stage model, bird information was masked to help the model generalize well on birds absent in train soundscapes. 
Finally, we further improve our performance by using a series of \textit{False-Positives} and \textit{False-Negatives} reduction techniques and the use of two different thresholds for bird call or ''nocall`` identification and bird categorization.
We reduced the false negatives by increasing the confidence of most frequent birds from each site by 0.1. This strategy worked well both on CV and public/private LB.
Therefore, there were three types of predictions, (a) only birds, (b) only nocall (no birds), and (c) both nocall and birds.

\subsubsection{Second-Stage Model}
The CNN model we trained had some limitations. The short train audio was labelled only at the clip level. On analysing train soundscapes, we found that if a bird is found anywhere in the clip, it increases the chances of finding the same bird at other places in the clip. Furthermore, the chances of finding the birds in immediate neighbour frames would also be high. This phenomenon encouraged us to train a second stage model on train soundscapes which calibrates the confidence using some frame-based and clip-based features. Some of the challenges in training the second stage model involved the limited number of birds and sites in train soundscapes, which could hurt the model’s generalisation capability. To solve this, we converted the multi-label problem into a binary classification problem and masked the information about birds for this second stage model. We started with training a simple logistic regression model where each unique tuple of (clip, 5-second frame, bird) constitutes a single training sample. No information about bird class was passed in any way directly to the model. This post-processing alone gave a $0.005$-$0.007$ boost on CV and public/private LB. We saw further improvement ($+0.002$) by adding location-based features and switching from logistic regression to support vector machines.
Only four features were used for training the second-stage model. For example, let us denote the probability generated for a 5-second frame ending at $k$ seconds for any bird B in the clip by $P_k$. Let the length of the clip be $n$ seconds ($n=600$ for all soundscapes). The calculation details of these features for the frame ending at $k$ seconds for the bird B are explained below:
\begin{enumerate}
    \item Frame-based features: Rolling Mean 3 ($RM_3$) and Rolling Mean 9 ($RM_9$)\\
    \[RM_{3} = \frac{1}{3}\sum_{i=k-1}^{k+1}P_{i}\]
    \[RM_{9} = \frac{1}{9}\sum_{i=k-4}^{k+4}P_{i}\]
    \item  Clip-based features:\\
    \[Maximum \; Confidence = max\, (P_{5} , \dots , P_{n})\]
    \item Distance-based features (minimum Haversine distance) explained in Section \ref{sub:min_dist}.
    
\end{enumerate}

\subsubsection{Minimum Haversine Distance}
\label{sub:min_dist}
The haversine distance \cite{haversine} is an excellent approximation for the angular distance between two points expressed as latitudes and longitudes on earth. The minimum haversine distance is expressed as the distance between a site and a bird class. Let us suppose that a bird class has 400 samples in the train short audios. First, the haversine distance is calculated between the position of each of those birds and the site’s location. The minimum of the set of these 400 distances is called minimum haversine distance.

\subsubsection{False Positives Reduction}
All the (bird, site) pairs satisfying at least one of the following conditions were discarded:
\begin{enumerate}
    \item Minimum Haversine Distance between site and bird is greater than 100. Analysing train soundscapes, we found that only 3 (birds, site) pairs found in train soundscapes have a minimum haversine distance greater than 60. So, all the (birds,site) pairs with minimum haversine distance>100 were rejected.
    \item Probabilities generated directly from the ensemble for that frame were less than 0.01. This post processing helped us get  small boost in CV and private LB and helped us reduced training data for Support Vector Classifier.
    \item  Remove birds belonging to one of the following classes - (Great Horned Owl, Plumbeous Pigeon). As we analysed the train soundscapes, we found that our models have a very high False Positives Rate for these species. Most of the time, when the model was predicting these classes, the actual target was nocall. 
\end{enumerate}


\subsubsection{Confidence Thresholds}
Two sets of thresholds were used for calibrating confidence. The nocall confidence was determined as $1 - max(\text{calibrated confidence for all birds for that 5-second frame})$. The first threshold was applied on nocall confidence (no birds detected). All 5-second frames having nocall confidence above this threshold contained nocall as one of the predictions. The second threshold was applied on calibrated confidence for each bird.

\section{Experimental Results}
\label{sec:exp}

\subsection{Evaluation of Methods}
We trained models using short train audios as explained in Section \ref{sec:pre}. We use long soundscape audios for training probability calibration (PC) model, for optimizing false-negative reduction (FNR) and false-positive reduction (FPR) methods, tuning thresholds, and for computing the validation scores (see Section \ref{sec:infer}). For \textbf{model selection}, we kept track of both call and nocall F1 scores to make sure that models are not heavily affected distribution of nocall-call samples. Note that Train soundscapes had around 63\% nocall samples, and we estimate that the hidden test fraction corresponding to the public LB has 54\% nocall samples.
We rely on two different validation scores: the ''High nocall Validation Score`` denoted as HNVS and the ''Low nocall Validation Score`` denoted as LNVS. In Equation \ref{metrics} we show the definition of both metrics:

\begin{equation}
\begin{aligned}
    \text{HNVS (CV@0.63)} = 0.63 \times \text{F1-micro}_{\text{nocall}} + 0.37 \times \text{F1-micro}_{\text{call}} \\
    \text{LNVS (CV@0.54)} = 0.54 \times \text{F1-micro}_{\text{nocall}} + 0.46 \times \text{F1-micro}_{\text{call}}
\end{aligned}
\label{metrics}
\end{equation}

In order to make sure that our models generalizes to unseen data (e.g. private LB), we separately calculated row-wise micro averaged F1-score for samples having bird calls and samples having no bird call. 
Table \ref{table:cv_lb_ensemble_models} shows the different metrics that we considered for selecting our models and experiments and the ablation study of the different postprocess steps. 
For further validation, we also tested our models on the Cornell Birdcall Identification 2020 Challenge leaderboard \footnote{\url{https://www.kaggle.com/c/birdsong-recognition/leaderboard}}. Last year competition data had 3 different sites,
after some analysis, we found that site2 was close to SSW site and site1 was close to SNE site.
Also we estimated that this test data has around 57\% nocall samples.

\subsection{Results and Comparison}

Table \ref{table:cv_lb_ensemble_models} summarizes our experiments. We bagged 13 CNN-based models (Section \ref{sec:models}) with CV@0.63 (HNVS) varying from 0.68 to 0.71. These 13 models were different in terms of augmentation strategy and architecture. Adding augmentations improved the true positive rate of these models and reduced the difference of scores between the predictions on short audios and train soundscapes(relatively noisier), thus making models robust against the anthropogenic noise. The bagging of these 13 models gave 0.74 CV@0.54 for COR site but was not that effective on SSW sites. For SNE \& SSW sites, we fine-tuned the last year competition first and second place models (only for birds having minimum Haversine distance lesser than 100 for these two sites). Using these models improved the CV@0.54 for SNE \& SSW sites from 0.69 to 0.75.

\begin{table*}[ht]
\centering
\caption{Row-wise micro averaged F1-score results of models on Public-Private LB, and ''train soundscape`` validation. For local validation, row-wise micro averaged F1-score was calculated on samples with call and no\_call separately, and the metric CV@0.54 (LNVS) was also calculated.}
\resizebox{\textwidth}{!}{
    \begin{tabular}{lccccccccccc}
        \toprule
        Method & \multicolumn{2}{c}{All Sites (2021)} & \multicolumn{3}{c}{COR Site} & \multicolumn{3}{c}{SSW Site} & \multicolumn{3}{c}{COR \& SSW Sites} \\
        & Public LB & Private LB & No call & Call & CV@0.54 & No call & Call & CV@0.54 & No call & Call & CV@0.54 \\
        \midrule
        SNE \& SSW site models & - & - & - & - & - & 0.9094 & 0.5552 & 0.7465 & - & - & -\\
        All site models & 0.7155 & 0.6203 & 0.9300 & 0.5208 & 0.7418 & 0.9431 & 0.3876 & 0.6875 & 0.9261 & 0.4623 & 0.7127 \\
        Ensemble & 0.7499 & 0.6450 & 0.9300 & 0.5208 & 0.7418 & 0.8923 & 0.5861 & 0.7514 & 0.9130 & 0.5591 & 0.7502 \\
        Ensemble + PC & 0.7744 & 0.6609 & 0.9187 & 0.6415 & 0.7912 & 0.8869 & 0.6106 & 0.7598 & 0.9044 & 0.6234 & 0.7751 \\
        Ensemble + PC + Site-info & 0.7711 & 0.6722 & 0.9106 & 0.6756 & 0.8025 & 0.8725 & 0.6327 & 0.7622 & 0.8934 & 0.6505 & 0.7816 \\
        Ensemble + PC + FNR & 0.7774 & 0.6630 & 0.9086 & 0.6758 & 0.8015 & 0.8720 & 0.6354 & 0.7632 & 0.8921 & 0.6521 & 0.7817 \\
        Ensemble + PC + FNR + FPR & 0.7754  & 0.6780 & 0.9285 & 0.6583 & 0.8029 & 0.8836 & 0.6343 & 0.7656 & 0.9082 & 0.6443 & 0.7836 \\
        \rowcolor{LightCyan} \textbf{Selected Submission} & \textbf{0.7801}  & \textbf{0.6738} & 0.9106 & 0.6756 & 0.8025 & 0.8754 & 0.6363 & 0.7654 & 0.8947 & 0.6526 & 0.7834 \\
        \midrule
    \end{tabular}}
    \label{table:cv_lb_ensemble_models}
\end{table*}

As shown in Table \ref{table:cv_lb_ensemble_models}, the time series-based probability calibration (PC) model provided a good improvement in CV and LB. Using the clip-level and the neighboring frames information, the probability calibration model improved CV on samples having bird call by +0.07, and raised Public LB to 0.774 and Private LB to 0.661.
Then the bird-to-site mapping (site-info) using minimum Haversine distance helped in two ways: (i) reducing false-negatives by reducing the call identification thresholds of the most frequent birds, (ii) reducing false positives by removing the birds in the predictions which are not found at a particular site.\\ 
Further reducing false negatives via removing birds from the predictions which are most commonly confused with ''nocall`` helped in achieving CV@0.54=0.7836 and Private LB=0.6780.
Additionally, we compare our current solution against previous state-of-the-art methods for this challenge by submitting our solution to Cornell Birdcall Identification 2020 Challenge.
Our solution was able to give significantly better results than previous winning solutions of the Cornell Birdcall Identification 2020 Challenge (see Table \ref{table:lb_2020}).
There is a gain of $+0.035$ on Public LB and a gain of $+0.018$ on Private LB as compared to 2020 first place solution.
We understand that our model is an extension of previous state of the art, and can generalize to detect all variety of birds from unknown sites and background sounds.

\begin{figure}[]
    \centering
    \includegraphics[width=\textwidth]{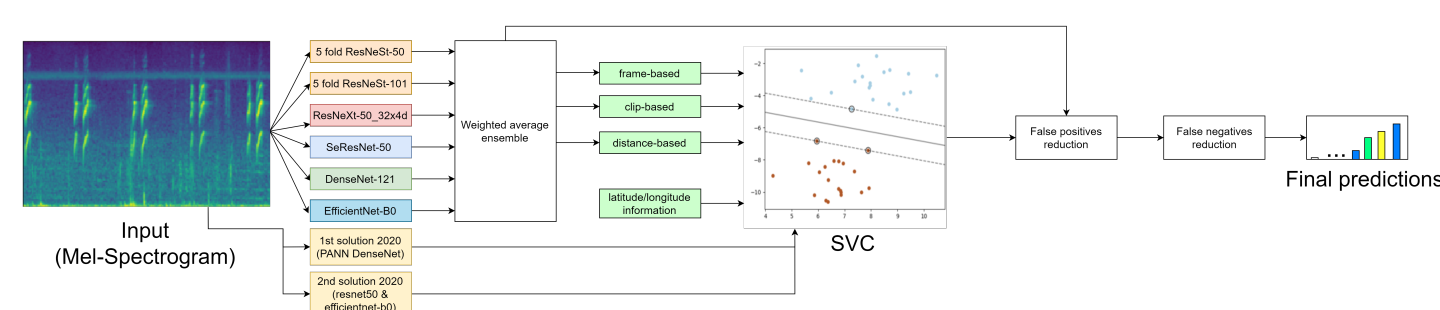}
    \caption{Overview of our solution pipeline. We show an ensemble of various models, our SVC model for probability calibration (PC) and the proposed probability filters for False-Positives and False-Negatives reduction.}
    \label{fig:ensemble}
\end{figure}

\begin{table}[]
    \centering
    \caption{Comparison of our solution ot the BirdCLEF 2021 Challenge (see Section \ref{sec:method}) and the winning solutions of the Cornell Birdcall Identification 2020 competition on its Leaderboard (public and private). Our solution generalizes to different sites and extends previous approaches improving performance.}
    \begin{tabular}{lcc}
        \toprule
        Model & \multicolumn{2}{c}{Cornell Birdcall Identification 2020}\\
        & Public LB & Private LB \\
        \midrule
        Ours (BirdCLEF 2021) & \textbf{0.659} & \textbf{0.699} \\
        Birdcall 2020 - 1st place & 0.624 & 0.681 \\
        Birdcall 2020 - 2nd place & 0.628 & 0.677 \\
        Birdcall 2020 - 3rd place & 0.626 & 0.675 \\
        \bottomrule
    \end{tabular}
    \label{table:lb_2020}
\end{table}

\section{Conclusion and Future work}
\label{sec:conclusion}

We aim to help researchers monitoring birds and automatically intuit factors about an area’s quality of life, levels of pollution, and the effectiveness of restoration efforts. We present a sound detection and classification pipeline for analyzing soundscape recordings that learns from weak labels, classifies fine-grained bird vocalizations and is robust against anthropogenic or natural noisy sounds (e.g., rain, cars, etc). Our solution achieved 10th place of 816 teams at the BirdCLEF 2021 Challenge.
We would like to improve efficiency and usability, and thus, use this pipeline online or on smartphones. To achieve this, we are exploring Knowledge Distillation to reduce notably the hardware requirements and inference time.

\begin{acknowledgments}
We would like to thank Kaggle and Dr.~Stefan Kahl for hosting the BirdCLEF 2021 Challenge. We also want to thank participants of the Cornell Birdcall Identification 2020 Challenge and this challenge for sharing insights, datasets, their solutions and open-sourced code, especially: Ryan Wong, Kramarenko Vladislav, Hidehisa Arai, Kossi Neroma, Jean-François Puget (CPMP).

\end{acknowledgments}

\bibliography{paper}


\end{document}